\documentclass[extra]{gji2}
\usepackage{graphicx}
\usepackage{amsmath,amssymb}
\usepackage{hyperref}
\usepackage{lineno}

\pdfoutput=1
\title[]{Influence of fluids on \vpvs{} ratio: Increase or decrease?}

\author[N. Brantut, E.C. David]{Nicolas Brantut and Emmanuel C. David\\Department of Earth Sciences\\ University College London, London, UK}

\date{}

\newcommand\vpvs{$V_\mathrm{P}/V_\mathrm{S}$}
\newcommand\nuc{$\nu_{0,\mathrm{crit}}$}
\newcommand\natexlab[1]{#1}

\begin{document}
\maketitle

\begin{summary}
  The evolution of the ratio between P- and S-wave velocity (\vpvs{}) with increasing fluid-saturated porosity is computed for isotropic rocks containing spheroidal pores. The ratio \vpvs{} is shown to either decrease or increase with increasing porosity, depending on the aspect ratio $\alpha$ of the pores, fluid to solid bulk modulus ratio $\zeta$, and Poisson's ratio $\nu_0$ of the
 solid constituents of the rock. A critical initial Poisson's ratio \nuc{} is computed, separating cases where \vpvs{} increases (if $\nu_0<\nu_\mathrm{0,crit}$) or \emph{decreases} (if $\nu_0>\nu_\mathrm{0,crit}$) with increasing porosity. For thin cracks and highly compressible fluids, \nuc{} is approximated by $0.157\,\zeta/\alpha$, whereas for spherical pores \nuc{} is given by $0.2 + 0.8\zeta$. When $\nu_0$ is close to \nuc{}, the evolution of \vpvs{} with increasing fluid-saturated porosity is near neutral and depends on subtle changes in pore shape and fluid properties. This regime is found to be relevant to partially dehydrated serpentinites in subduction zones (porosity of aspect ratio near 0.1 and $\zeta$ in the range 0.01--0.1), and makes detection of these rocks and possibly elevated fluid pressures difficult from \vpvs{} only.
\end{summary}

\section{Introduction}



The ratio of P- to S-wave velocities (\vpvs{} ratio) is commonly considered as a key constraint on the nature and composition of rocks when interpreting seismological data \citep[e.g.,][]{christensen96}. It is also well established that the presence of fluid-filled porosity (cracks, pores, or open grain junctions) strongly modifies the \vpvs~ratio \citep[e.g.,][among many others]{oconnell74,kuster74,watanabe93,zimmerman94,leravalec96,berryman02,takei02,fortin07}. 
%
Two systematic observations are that (1) full saturation leads to an increase in \vpvs{} compared to dry rocks \citep[e.g.][]{nur69}, and (2) the opening of \emph{liquid-saturated cracks} (e.g., when confining pressure is reduced) also causes an increase in \vpvs{} \citep[see experimental data by][]{christensen84}. Both observations are well supported by theoretical models based on effective medium schemes in cracked materials \citep[e.g.,][]{oconnell74,berryman02}, and have been used to interpret results from seismic tomography \citep[e.g.,][]{peacock11}. 
However, when the fluid compressibility is very large compared to that of the rock, or when the fluid is present in pores shaped differently from thin cracks (e.g., tubes, spherical pores or polygonal grain-junctions), the change in \vpvs{}  with increasing fluid-saturated porosity is not necessarily a monotonic increase. For instance, using an effective medium theory based on fluid inclusions in the shape of triangular tubes, \citet{watanabe93} showed that, with increasing porosity, \vpvs{} initially decreases if the porosity is saturated with water, whereas it increases if the porosity is saturated with a much less compressible fluid such as melt. Similarly, the comprehensive review presented by \citet{takei02} shows that  a regime exists where \vpvs{} decreases with increasing fluid content, notably for gas-saturated cracks \citep[see also][]{dvorkin99} and for texturally equilibrated water-saturated inclusions.

Overall, models based on effective medium approaches show that pore geometry and fluid compressibility have a strong influence on the variations in \vpvs{} (or, equivalently, Poisson's ratio) with increasing fluid content \citep[e.g.,][]{oconnell74,zimmerman94,berryman02}. However, an important control parameter that has been apparently overlooked is the Poisson's ratio of the host material. In most modelling studies, it is taken equal to 0.25 for simplicity, and a systematic exploration of this parameter has rarely been undertaken \citep[a notable exception is][for the case of spherical inclusions]{zimmerman91b,zimmerman94}. In addition, published models often require systematic computations of bulk and shear moduli as a function of fluid-saturated porosity to access the evolution in \vpvs{}, but it is desirable to achieve approximate predictions using simple formulae that exhibit clearly how the three key parameters (Poisson's ratio of solid constituents, pore shape and fluid compressibility) influence the results.

Here, we use the differential effective medium scheme to determine the variations of \vpvs{} in materials containing an isotropic distribution of fluid-filled spheroidal inclusions. We determine the critical parameter values separating cases when \vpvs{} increases or decreases with increasing porosity, and provide simple closed-form asymptotes for limiting pore shapes (cracks, spheres and needle-like cavities). Finally, we discuss a number of geophysically relevant cases where the presence of fluids may have a counterintuitive impact on \vpvs{}.



\section{Methodology}

We use (1) the differential effective medium (DEM) scheme to compute the effective elastic properties of solids containing voids (i.e., dry pores), and (2) the Gassmann relationship to compute the effect of a fluid filling the voids. Some of our analysis is given in the limit of small porosity (see Section \ref{sec:nuc}) and is therefore general and does not rely specifically on the DEM approximation \citep{zimmerman91}.

The DEM approach consists in incrementally introducing inclusions (amounting to an increment of porosity), computing the corresponding incremental change in effective elastic moduli, and repeating the procedure until the target porosity is reached \citep[e.g.,][]{bruner76,mclaughlin77,henyey82,zimmerman84,norris85}. For an isotropic solid containing randomly oriented, spheroidal voids of a given aspect ratio $\alpha$, the effective bulk ($K$) and shear ($G$) moduli are given by the following set of coupled ordinary differential equations \citep[e.g.,][]{david12b}
\begin{linenomath}
  \begin{align}
    \frac{1-\phi}{K}\frac{dK}{d\phi} &= -P(\alpha, \nu),\label{eq:dKdphi}\\
    \frac{1-\phi}{G}\frac{dG}{d\phi} &= -Q(\alpha, \nu),\label{eq:dGdphi}
  \end{align}
\end{linenomath}
where $\phi$ is the porosity and $\nu$ is Poisson's ratio of the effective dry porous material. Poisson's ratio is related to the elastic moduli as
\begin{linenomath}
  \begin{equation} \label{eq:nu}
    \nu = \frac{3K-2G}{6K+2G}.
  \end{equation}
\end{linenomath}
The elastic constants of the intact material (at $\phi=0$), i.e., of the solid constituents of the rock matrix, are denoted $K_0$, $G_0$ and $\nu_0$. The functions $P$ and $Q$ are the bulk and shear compliances of the spheroidal void, respectively, and depend on the Poisson's ratio $\nu$ of the dry porous solid, and the aspect ratio $\alpha$ of the spheroids. Full expressions for $P$ and $Q$ are given in \citep{david11}. In using expressions \eqref{eq:dKdphi} and \eqref{eq:dGdphi}, we assume the existence of a unique family of pores of the same representative aspect ratio $\alpha$, and use the porosity $\phi$ as our control parameter. More complex microstructures could be represented by using a combination of pores of different aspect ratios, for instance mixtures of thin cracks and spherical pores, and using specific concentration parameters for each family (e.g., crack density and porosity, see \citet{shafiro97}). Here, we restrict our attention to a single aspect ratio in order to highlight the controlling role of this parameter and keep the analysis as simple as possible. 

From the effective moduli of the dry porous material, the moduli of the fluid-saturated material are given by Gassmann's fluid-substitution relations in the undrained limit \citep{gassmann51}:
\begin{linenomath}
  \begin{align} 
    K_\mathrm{u} &\displaystyle=  K\frac{\phi(1-\zeta^{-1}) + 1-K_0/K}{\phi(1-\zeta^{-1})+K/K_0-1}, \label{eq:gassmannK}\\
    G_\mathrm{u} &= G,\label{eq:gassmannG}
  \end{align}
\end{linenomath}
where subscripts $\mathrm{u}$ indicate saturated moduli, $K$ is the dry effective bulk modulus, and
\begin{linenomath}
  \begin{equation} \label{eq:zeta}
    \zeta = K_\mathrm{f}/K_0
  \end{equation}
\end{linenomath}
is the ratio of the bulk moduli of the fluid and of the solid constituents of the material. From the elastic moduli, we compute Poisson's ratio using \eqref{eq:nu} and the \vpvs{} ratio as
\begin{linenomath}
  \begin{equation}
    \frac{V_\mathrm{P}}{V_\mathrm{S}} = \sqrt{\frac{2(1-\nu)}{1-2\nu}}.
  \end{equation}
\end{linenomath}

\section{Dry limit}


Before investigating the effect of fluids \textit{per se}, it is instructive to examine first the evolution of \vpvs{} with increasing \emph{dry} porosity. A full investigation was presented by \citet{david11b}, and only the key results are summarised here.

\begin{figure}
  \centering
  \includegraphics{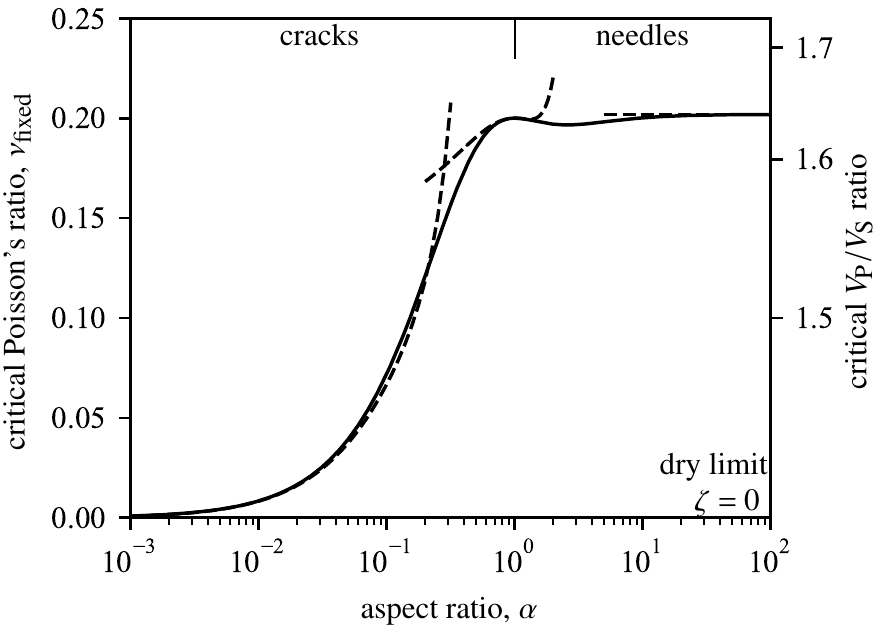}
  \caption{Fixed point for Poisson's ratio and \vpvs{} as $\phi\rightarrow 1$ in the dry case. Low aspect ratios $\alpha<1$ correspond to oblate spheroids (crack-like shapes), high aspect ratios $\alpha>1$ correspond to prolate spheroids (needle-like shapes). Spheres correspond to $\alpha=1$. The solid line is the numerical solution, and dashed lines are the closed-form asymptotes for thin cracks, spheres, and needles.}
  \label{fig:dry}
\end{figure}

Combining Equations \eqref{eq:nu}, \eqref{eq:dKdphi} and \eqref{eq:dGdphi}, an ordinary differential equation for Poisson's ratio is obtained:
\begin{linenomath}
  \begin{equation} \label{eq:dnudphi}
    (1-\phi)\frac{d\nu}{d\phi} = \frac{(1+\nu)(1-2\nu)}{3}\big[Q(\alpha,\nu)-P(\alpha,\nu)\big].
  \end{equation}
\end{linenomath}
As shown in \citet{berryman02} and \citet{david11b}, Poisson's ratio evolves monotonically with increasing porosity towards a fixed point $\nu_\mathrm{fixed}$ (where $d\nu/d\phi=0$) that depends only on the aspect ratio of the pores, and is independent from the moduli of the solid constituents of the material. Qualitatively similar results hold for other effective medium schemes \citep[e.g.,][]{dunn95}. The fixed point $\nu_\mathrm{fixed}$ can be computed by setting $Q(\alpha,\nu_\mathrm{fixed})=P(\alpha,\nu_\mathrm{fixed})$, and is shown in Figure \ref{fig:dry} (solid line). Closed-form solutions in asymptotic cases (Appendix \ref{ax:dry}) are obtained for
\begin{description}
\item[thin cracks $(\alpha\ll1)$:]
  \begin{linenomath}
    \begin{equation}
      \nu_\mathrm{fixed} \simeq 0.861\alpha - 2.504\alpha^2 + 5.882\alpha^3, \label{eq:nufixed_crack}
    \end{equation}
  \end{linenomath}
\item[nearly spherical pores $(\alpha\sim1)$:]
  \begin{linenomath}
    \begin{equation}
      \nu_\mathrm{fixed} \simeq 0.200 - 0.018(1-\alpha)^2 - 0.039(1-\alpha)^3, \label{eq:nufixed_sphere}
    \end{equation}
  \end{linenomath}
\item[and needle-like pores $(\alpha\gg1)$:]
  \begin{linenomath}
    \begin{equation}
      \nu_\mathrm{fixed} \simeq 0.202. \label{eq:nufixed_needle}
    \end{equation}
  \end{linenomath}
\end{description}
These asymptotic solutions (Figure \ref{fig:dry}) show excellent agreement with the numerical solution over most of the aspect ratio range, except near the transition between thin cracks and spheres ($0.25\lesssim\alpha\lesssim0.6$) and between spheres and needles ($1.4\lesssim\alpha\lesssim11$). The asymptote for thin cracks \eqref{eq:nufixed_crack} differs from that of \citet{berryman02} for penny-shaped cracks (their Equation B3) probably due to a typographical error \footnote{the asymptotic approximation given by \protect\citet{walsh69} for penny-shaped cracks --his Equation (1b)-- does not match the one rederived in \protect\citet{david11} in the dry case. Reconciling the two expression requires removing the first unitary term on the right-hand side of Walsh's Equation (1b)} propagated in the literature since \citet{walsh69} (reproduced notably in \citet{berryman80}).

The key result of the analysis, illustrated in Figure \ref{fig:dry}, is that $\nu_\mathrm{fixed}$ acts as a critical boundary separating materials (and pore shapes) for which increasing porosity produces a decrease or increase in $\nu$ (and \vpvs{}). When the Poisson's ratio of the intact material (i.e., that of the solid constituents of the rock) $\nu_0$ is greater than $\nu_\mathrm{fixed}(\alpha)$, then $\nu$ \textit{decreases} with increasing porosity. In the example of thin cracks, $\nu_\mathrm{fixed}\rightarrow0$ as $\alpha\rightarrow0$, so that Poisson's ratio is systematically decreasing with increasing crack porosity. Such a behaviour is confirmed by laboratory experiments on gas-saturated (or dry) cracked rocks \citep[e.g.,][]{dvorkin99}. Interestingly, \textit{any} material for which $\nu_0\gtrsim0.2$ will see its Poisson's ratio decrease with increasing (dry) porosity, regardless of pore shape.

\section{Compressible fluids}

\subsection{General DEM results}

At a given porosity, $\nu$ and \vpvs{} are always higher for the fluid-saturated porous solid than for its dry counterpart. This result is independent of the pore shape, and is a direct consequence of (1) the increase of $K/G$ in the presence of fluids (Gassmann's equation, \eqref{eq:gassmannK} and \eqref{eq:gassmannG}), and (2) $\nu$ being an increasing function of $K/G$.

When the porosity is saturated with a compressible fluid, the evolution in $\nu$ and \vpvs{} with increasing porosity differs significantly from the dry case. From a physical point of view, one expects that $\nu$ should tend to $0.5$ ($V_\mathrm{P}/V_\mathrm{S}\rightarrow+\infty$) as $\phi\rightarrow1$ (i.e., when the material is effectively just a fluid). For a saturating fluid of low compressibility, one also expects that $\nu$ should closely follow the evolution in the dry case at low porosity, before transitioning to an eventual increase towards $0.5$ at high porosity. Such basic physical arguments indicate that the evolution of $\nu$ with porosity might be complex and non-monotonic.

\begin{figure*}
  \centering
  \includegraphics{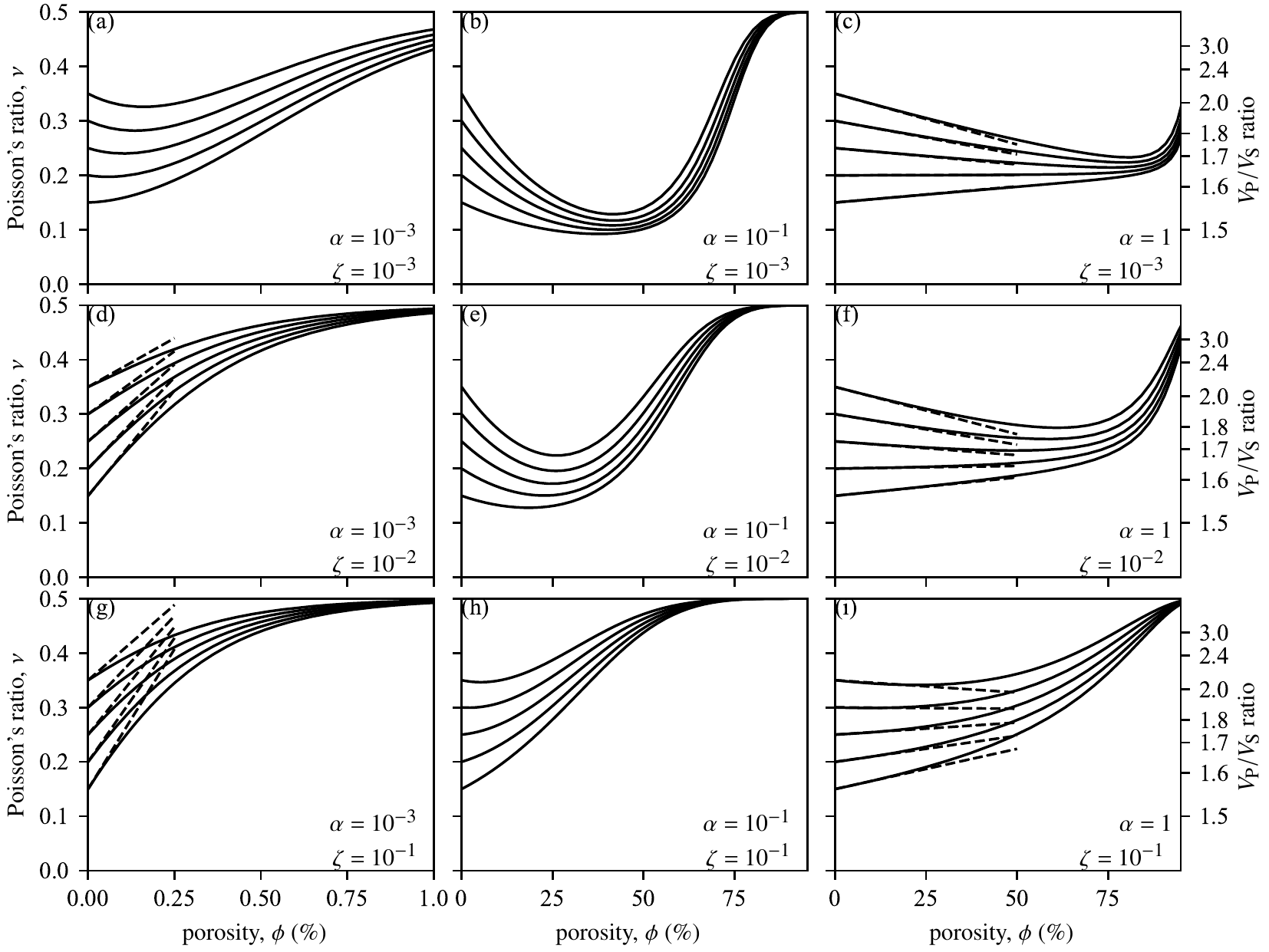}
  \caption{Evolution of Poisson's ratio and \vpvs{} with increasing fluid-saturated porosity for a range of pore aspect ratios ($\alpha=10^{-3}$, $10^{-1}$ and $1$) and fluid compressibility ratios ($\zeta=K_\mathrm{f}/K_0=10^{-3}$, $10^{-2}$ and $10^{-1}$). Solid lines are numerical solutions to the DEM and Gassmann's Equations (Equations \eqref{eq:dKdphi}, \eqref{eq:dGdphi} \eqref{eq:gassmannK} and \eqref{eq:gassmannG}) for initial Poisson's ratio ranging from $\nu_0=0.15$ to $\nu_0=0.35$. Dashed lines are asymptotic solutions obtained to first order in $\phi$ for thin cracks ($\alpha\ll\zeta\ll1$) and spheres ($\alpha=1$).}
  \label{fig:all_undrained}
\end{figure*}

Complete numerical solutions for the DEM combined with Gassmann's relationship are shown in Figure \ref{fig:all_undrained}. Each panel of Figure \ref{fig:all_undrained} shows results for $\nu$ and \vpvs{} as function of porosity at fixed $(\alpha, \zeta)$ and for $\nu_0=0.15,\,0.20,\,0.25,\,0.30$ and $0.35$. For thin cracks ($\alpha=10^{-3}$, left panels), results are only shown for $\phi$ up to 1\%.

For thin cracks filled with a low compressibility fluid ($\alpha=10^{-3}$, $\zeta=10^{-2}$ and $\zeta=10^{-1}$, Figures \ref{fig:all_undrained}d,g), Poisson's ratio increases rapidly towards $0.5$ as porosity increases to around 1\%, regardless of the solid's Poisson's ratio $\nu_0$. For $\alpha=10^{-3}$ and $\zeta=10^{-3}$ (Figure \ref{fig:all_undrained}a), $\nu$ also rises rapidly to $0.5$ at $\phi>0.2$\%, but the initial evolution $\nu(\phi)$ depends on $\nu_0$. For $\nu_0=0.15$, $\nu$ increases monotonically with increasing porosity. For $\nu_0\geq0.2$, the evolution at small porosity is a decrease in $\nu$, followed by an increase at $\phi\geq0.2$\%.

In the case of spherical pores ($\alpha=1$, Figures \ref{fig:all_undrained}c,f,i), the evolution of Poisson's ratio with increasing porosity depends on its initial value $\nu_0$. For $\phi<50$\%, $\nu$ increases if $\nu_0$ is less than around $0.2$, and decreases otherwise. At some large critical porosity (that depends on $\zeta$), $\nu$ rapidly increases to $0.5$. The strong variation of $\nu$ at $\phi$ near 100\% has been discussed in detail by \citet{zimmerman94}.

At intermediate aspect ratios ($\alpha=0.1$, Figures \ref{fig:all_undrained}b,e,h), $\nu$ typically evolves non-monotonically with increasing porosity. At low $\zeta$ ($\zeta\leq 10^{-2}$, highly compressible fluids), $\nu$ tends to decrease with increasing $\phi$ up to $\phi\approx40$\% (at $\zeta=10^{-3}$) and $\phi\approx20$\% (at $\zeta=10^{-2}$), before eventually increasing towards $0.5$. For less compressible fluids ($\zeta=10^{-1}$), the initial evolution of $\nu$ strongly depends on $\nu_0$: for $\nu_0=0.35$, $\nu(\phi)$ initially decreases, whereas it is either stable or increases at $\nu_0\leq0.3$.


\begin{figure*}
  \centering
  \includegraphics{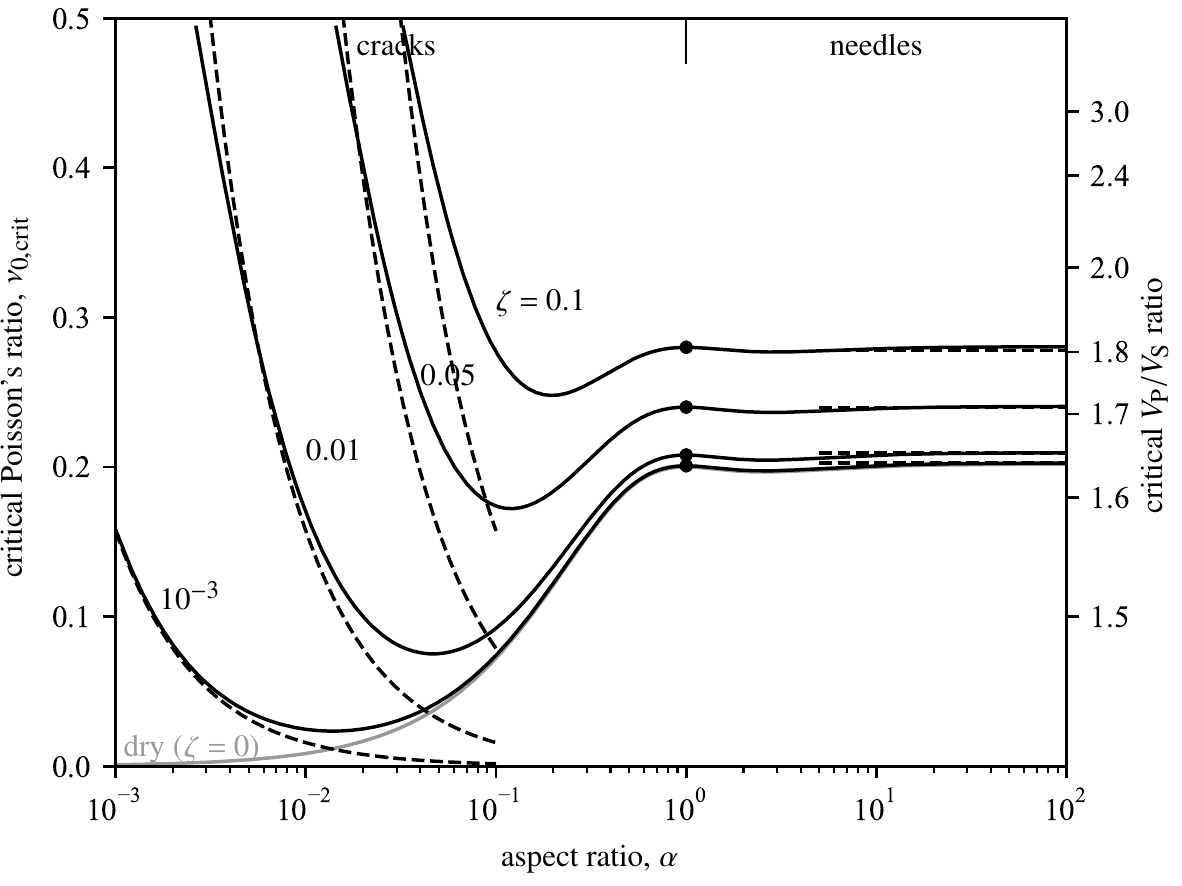}
  \caption{Critical initial Poisson's ratio and \vpvs{} separating increasing or decreasing $\nu(\phi)$ at $\phi=0$. For $\nu_0>\nu_\mathrm{0,crit}$, $\nu$ (and \vpvs{}) initially decreases with increasing fluid-saturated porosity. Solid lines are numerical solutions. Dashed lines and black circles are asymptotic closed-form expressions for thin cracks, needle-like pores and spherical pores, respectively. The solid grey curve corresponds to the dry case (same as in Figure \ref{fig:dry}).}
  \label{fig:nucrit_undrained}
\end{figure*}


\subsection{Critical parameters separating increase from decrease in $V_\mathrm{P}/V_\mathrm{S}$} \label{sec:nuc}

One way to understand the numerical results from the DEM approach is to determine the critical parameter values separating the cases where $d\nu/d\phi<0$ and $d\nu/d\phi>0$ at small $\phi$, i.e., at the introduction of fluid-saturated pores in the solid. We define a critical initial Poisson's ratio $\nu_{0,\mathrm{crit}}(\alpha, \zeta)$ such that
\begin{linenomath}
  \begin{equation} \label{eq:nu0critdef}
    \text{if} \quad \nu_0>\nu_{0,\mathrm{crit}}\quad\text{then}\quad\frac{d\nu}{d\phi}\Big|_{\phi=0}<0.
  \end{equation}
\end{linenomath}
Since \nuc{} is defined in the limit $\phi\rightarrow0$, the following analysis is not specific to the DEM approximation (when $\phi\rightarrow 0$, all effective medium schemes produce the same predictions).




The qualitative evolution of \nuc{} with increasing aspect ratio is similar for all tested values of $\zeta$. At low $\alpha$, \nuc{} initially decreases with increasing aspect ratio, and then increases up to a plateau at $\alpha\geq1$. The transition point where \nuc{} is minimum scales with the ratio $\zeta/\alpha$. The value of \nuc{} at aspect ratios above 1 depends on $\zeta$ but not significantly on $\alpha$. For $\zeta\ll\alpha\ll1$, the evolution of \nuc{} closely follows that of $\nu_\mathrm{fixed}$ in the dry case.

Asymptotic expressions for \nuc{} can be determined in simple cases (see Appendix \ref{ax:undrained}):
\begin{description}
\item[thin cracks $(\alpha\ll\zeta\ll1)$:]
  \begin{linenomath}
    \begin{equation}
      \nu_\mathrm{0,crit} \simeq 0.157\frac{\zeta}{\alpha},\label{eq:nucrit_crack}
    \end{equation}
  \end{linenomath}
\item[spheres $(\alpha\sim1,\,\zeta\ll1)$:]
  \begin{linenomath}
    \begin{equation}
      \nu_\mathrm{0,crit} \simeq 0.2 + 0.8\zeta, \label{eq:nucrit_sphere}
    \end{equation}
  \end{linenomath}
\item[and needles $(\alpha\gg1,\,\zeta\ll1)$:]
  \begin{linenomath}
    \begin{equation}
      \nu_\mathrm{0,crit} \simeq 0.202 + 0.760\zeta. \label{eq:nucrit_needle}
    \end{equation}
  \end{linenomath}
\end{description}
The accuracy of these approximations is excellent at very low $\zeta$, but deteriorates with increasing $\zeta$, especially in the case of thin cracks (Figure \ref{fig:nucrit_undrained}). More accurate asymptotes could probably be determined with higher order expansions in terms of $\alpha$ and $\zeta$, but we retain formulae \eqref{eq:nucrit_crack}, \eqref{eq:nucrit_sphere} and \eqref{eq:nucrit_needle} because of their remarkable simplicity. For completeness, Appendix \ref{ax:unrelaxed} presents analogue asymptotes for the case of fluid-saturated rocks in the high-frequency (``unrelaxed'') limit, and shows only small or no quantitative differences with Equations \eqref{eq:nucrit_crack}, \eqref{eq:nucrit_sphere} and \eqref{eq:nucrit_needle}. The key result of our analysis, illustrated in Figure \ref{fig:nucrit_undrained}, is the prediction and elementary estimate for the critical Poisson's ratio of solid constituents of a rock above which the introduction of fluid-saturated pores produces a \emph{decrease} in the effective Poisson's ratio and \vpvs{}.

\subsection{Estimates of \vpvs{} at low porosity}

The comparison between Poisson's ratio of the solid constituents of the rock, $\nu_0$, and \nuc{}, provides a simple rule to predict whether fluid-filled porosity induces an increase or a decrease in the effective $\nu$ of the saturated porous rock. The amplitude of the variation of $\nu$ with $\phi$ ($d\nu/d\phi$ at $\phi=0$) is approximated by asymptotic expansions of the DEM and Gassmann's equations for
\begin{description}
\item[thin cracks $(\alpha\ll\zeta\ll1),\,\nu_0\lesssim0.25$:]
  \begin{linenomath}
    \begin{equation}
      \frac{d\nu}{d\phi}\Big|_{\phi=0} \sim \frac{20-34\nu_0}{45\pi\alpha} + \frac{1-\nu_0}{3}\left(1-\frac{1}{\zeta}\right),\label{eq:dnu_crack}
    \end{equation}
  \end{linenomath}
\item[spheres $(\alpha=1,\,\zeta\ll1)$:]
  \begin{linenomath}
    \begin{equation}
      \frac{d\nu}{d\phi}\Big|_{\phi=0} \sim \frac{3}{2}\frac{(1-5\nu_0)(1-\nu_0^2)}{7-5\nu_0} + \frac{3}{4}\frac{(1-\nu_0)^2(1+\nu_0)}{1-2\nu_0}\zeta, \label{eq:dnu_sphere}
    \end{equation}
  \end{linenomath}
\item[ and needles $(\alpha\gg1,\,\zeta\ll1)$:]
  \begin{linenomath}
    \begin{equation}
      \frac{d\nu}{d\phi}\Big|_{\phi=0} \sim \frac{(1+\nu_0)(5-28\nu_0+16\nu_0^2)}{15} + \frac{(1+\nu_0)(5-4\nu_0)^2}{27(1-2\nu_0)} \zeta. \label{eq:dnu_needle}
    \end{equation}
  \end{linenomath}
\end{description}
Linear approximation for $\nu(\phi)$ at small porosities using the above asymptotes are shown as dashed lines in Figure \ref{fig:all_undrained}. The approximation for spheres is remarkably accurate up to very large porosity ($\phi$ up to 25\%), while the approximation for thin cracks becomes poor at porosity larger than 0.1\%. More accurate, higher order asymptotes for the case of thin cracks could be obtained, but we only retain here the very approximate formula \eqref{eq:dnu_crack} for its simplicity, keeping in mind that full numerical solutions should be used at high crack porosity, large $\zeta$ and $\nu_0\gtrsim0.25$.






\section{Discussion}

\begin{figure}
  \centering
  \includegraphics{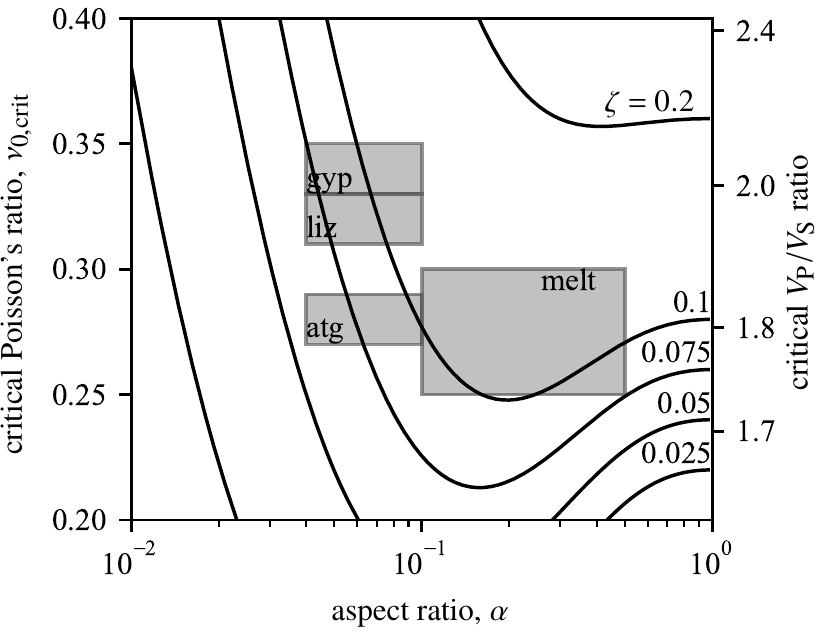}
  \caption{Critical Poisson's ratio and \vpvs{} as a function of aspect ratio and for indicated compressibility ratios $\zeta=0.02$--$0.2$. Grey boxes correspond to $(\alpha,\nu_0)$ ranges at the onset of gypsum dehydration (gyp), lizardite dehydration (liz), antigorite dehydration (atg) and silicate melting (melt).}
  \label{fig:nucrit_examples}
\end{figure}

Our modelling results demonstrate that the evolution in $\nu$ (or, equivalently, \vpvs{}) with increasing fluid-saturated porosity is potentially non-monotonic. The critical initial Poisson's ratio \nuc{} separating cases when \vpvs{} decreases or increases shows a complex evolution at pore aspect ratios near $\alpha=0.1$ and fluid compressibility ratios near $\zeta=0.1$. This range of parameters is typical of two key scenarios of geological relevance: metamorphic dehydration reactions and partial melting.

The laboratory experiments of \citet{popp93} and \citet{brantut12} showed that both serpentinite and gypsum undergoing thermal dehydration reactions see their Poisson's ratio \emph{decrease} with increasing reaction progress (i.e., with increasing fluid-saturated porosity). More specifically, \citet{brantut12} used the DEM approach to show that the pores generated by the transformation of gypsum to bassanite have an aspect ratio of the order of 0.05. This relatively large value is required due to the large porosity generated by dehydration reactions (typically of the order of 10\% or more), which cannot be accommodated by thin cracks only \citep{brantut12}. \citet{takei02} showed that equilibrated textures for partially molten and fluid-saturated rocks, where porosity is located at grain boundaries and triple junctions and is in equilibrium with surface tension forces, correspond to an effective material containing spheroidal pores of aspect ratio $\alpha=0.1$--$0.5$.

Using intact Poisson's ratio of $\nu_0=0.33-0.35$ for gypsum \citep{brantut12}, $\nu_0=0.31-0.33$ for lizardite \citep{popp93,christensen96}, and $\nu_0=0.26-0.28$ for antigorite \citep{reynard13}, the evolution of $\nu$ and \vpvs{} of these rocks at the onset of dehydration is in a regime where it is strongly controlled by the compressibility ratio $\zeta$ (Figure \ref{fig:nucrit_examples}). Gypsum dehydration occurs at low pressure and temperature, so that $K_\mathrm{f}\approx2$~GPa and $K_0\approx41$~GPa, yielding $\zeta\approx0.05$. This set of parameters is clearly in the regime where $\nu_0>\nu_\mathrm{0,crit}$, and Poisson's ratio is expected to decrease with increasing porosity, a prediction confirmed by experiments \citep{brantut12}. For the case of lizardite dehydration at around $400^\circ$C, the evolution of $\nu$ depends on the fluid pressure. Under the experimental conditions of the study by \citet{popp93}, the fluid pressure is expected to be commensurate to the confining pressure of $200$~MPa, so that the bulk modulus of water is of the order of $1$~GPa (at $400^\circ$C). Using $K_0\approx57$~GPa \citep[derived from][]{christensen96}, it is found that $\zeta\approx0.02$, well within the regime where $\nu_0>\nu_\mathrm{0,crit}$, so that $\nu$ decreases with increasing porosity, as confirmed by the experimental results. By contrast, if lizardite dehydration occurs at higher pressure, say $1$~GPa, the fluid bulk modulus is around $K_\mathrm{f}\approx5.5$~GPa, so that $\zeta\approx0.1$, and the resulting evolution of $\nu$ and \vpvs{} is neutral (at $\alpha$ near 0.1) or increasing (at $\alpha\lesssim0.07$). Similarly, the case of antigorite dehydration is also complex. At $1$~GPa pressure and $550^\circ$C, the bulk modulus of water is $K_\mathrm{f}\approx4.5$~GPa. Using a bulk modulus of $K_0\approx75$~GPa for pure antigorite \citep{bezacier13} results in $\zeta\approx0.06$, which places $\nu_0$ only slightly above \nuc. Therefore, antigorite dehydration is expected to produce \emph{constant} or slightly \emph{decreasing} $\nu$ and \vpvs{}.

By contrast, the case of partial melting of silicates is unambiguous. Using a lower crustal silicate melt compressibility in the range $18-27$~GPa \citep{stolper81} and silicate bulk modulus in the range $80-110$~GPa yields $\zeta\approx0.16-0.34$. For most silicate rocks, $\nu_0$ is in the range $0.2-0.3$, which is below the predicted \nuc{} for melt-saturated pores (Figure \ref{fig:nucrit_examples}), so that partial melting is expected to produce an increase in $\nu$ and \vpvs{}, in accordance with previous predictions by \citet{takei02}.

\section{Conclusions}

The results from the DEM approach demonstrate that the Poisson's ratio $\nu_0$ of the solid constituents of a rock exerts a key control on the evolution of $\nu$ and \vpvs{} with increasing fluid-saturated porosity. This control has often been overlooked and most modelling studies have instead focussed on the effect of pore shape and fluid compressibility, assuming $\nu_0=0.25$. Here, we computed a critical Poisson's ratio $\nu_\mathrm{0,crit}(\alpha, \zeta)$ separating the cases when $\nu$ (and \vpvs{}) \emph{decreases} (if $\nu_0>\nu_\mathrm{0,crit}$) or increases (if $\nu_0<\nu_\mathrm{0,crit}$) with increasing porosity. Our analysis of \nuc{} is given in the limit of small porosity, and is therefore independent from the choice of a specific effective medium scheme. Simple asymptotic formulae were derived in the case of thin cracks, spherical pores and needle-like pores (Equations \eqref{eq:nucrit_crack}, \eqref{eq:nucrit_sphere} and \eqref{eq:nucrit_needle}). When $\nu_0$ is very close to \nuc{}, the evolution of \vpvs{} with porosity is near neutral, but becomes sensitive to subtle changes in pore shape and fluid compressibility. This case is likely encountered during dehydration reactions of serpentinites, where the details of the pore shape (driven by textural equilibration of the microstructure) and fluid properties (which depend on the local pressure, temperature and chemical composition) are expected to drive \vpvs{} towards either a slight increase or a decrease. A significant decrease in \vpvs{} was observed during lizardite dehydration \citep{popp93}, in accordance to our model's prediction. More experimental work is needed to further test the model predictions over a wider range of conditions and materials.

We only treated the case of isotropic solids containing isotropic distributions of pore orientations. Anisotropic matrix or anisotropic pore orientation distributions are expected to change the expected \vpvs{} ratio which then depends on the polarisation of the seismic waves propagating through the material \citep{reynard10,wang12}. In natural scenarios, such as partially dehydrated rocks in subduction zones, the combined effects of initial rock properties, fluid properties, pore shape and anisotropy make structural interpretations difficult from the measurement of \vpvs{} only. Unambiguous identification of specific rock types (such as serpentinites) and locally elevated fluid pressures is therefore likely to require a combination of datasets, including wave speed anisotropy and attenuation.

\begin{acknowledgments}
  Robert Zimmerman is thanked for his major influence at the early stages of this work. This paper is dedicated to him. Comments and suggestions from Mark Kachanov, J\"org Renner and an anonymous reviewer helped clarify the paper. The UK Natural Environment Research Council supported this work through grants NE/K009656/1 to NB and NE/M016471/1 to NB and ECD. Codes are accessible at \url{https://www.github.com/nbrantut/poisson.git}.
\end{acknowledgments}


\appendix

\section{Asymptotic forms in the dry case}
\label{ax:dry}

The fixed point $\nu_\mathrm{fixed}$ is given by solving for $\nu$ in
\begin{linenomath}
  \begin{equation}\label{eq:nudry}
    Q(\alpha,\nu) - P(\alpha,\nu) = 0.
  \end{equation}
\end{linenomath}

\begin{description}
  \item[\textbf{Thin cracks} ($\alpha\ll1$)] The term $Q(\alpha,\nu) - P(\alpha,\nu)$ is a rational function of the variable $\nu$. Removing the unphysical root $\nu=1$, retaining the two dominant terms of order zero and one in $\nu$, and performing a Taylor expansion to third order in $\alpha$ yield the approximation
\begin{linenomath}
  \begin{align}
    \nu_\mathrm{fixed}&\sim\left(\frac{4}{3 \pi }+\frac{5 \pi }{36}\right) \alpha +\left(-\frac{254}{81}+\frac{80}{27 \pi ^2}+\frac{29 \pi^2}{864}\right) \alpha ^2 \nonumber\\
    &\qquad+ \frac{\left(1228800+1660160 \pi ^2+165504 \pi ^4+315 \pi ^6\right) }{186624 \pi ^3}\alpha ^3.
\end{align}
\end{linenomath}
\item[\textbf{Nearly spherical pores} ($\alpha\sim1$)] Taylor expansion of $P(\alpha,\nu)$ and $Q(\alpha,\nu)$ for $\epsilon=(1-\alpha)\sim0$ are used, and \eqref{eq:nudry} is then solved to yield a third-order approximation in $\epsilon$ as
\begin{linenomath}
  \begin{equation}
    \nu_\mathrm{fixed}\sim \frac{1}{5} -\frac{16 }{875}\epsilon ^2 + \frac{3017088 \left(5751377+23283 \sqrt{59385}\right) }{42875 \left(135+\sqrt{59385}\right)^4}\epsilon ^3.
  \end{equation}
\end{linenomath}
\item[\textbf{Needles} ($\alpha\gg1$)] The limits of $P$ and $Q$ for needles \citep[see][]{david11} are used, and the solution of \eqref{eq:nudry} gives
\begin{linenomath}
  \begin{equation}
    \nu_\mathrm{fixed}\sim \frac{1}{8}\left(7-\sqrt{29}\right),
  \end{equation}
\end{linenomath}
recovering the solution previously derived by \citet{berryman02}.
\end{description}




\section{Asymptotic forms in the saturated, undrained case}
\label{ax:undrained}


Here we only study the behaviour at small porosity, near $\phi=0$. The set of Equations \eqref{eq:dKdphi} and \eqref{eq:dGdphi} for the DEM scheme then reduce to the dilute approximation:
\begin{linenomath}
  \begin{align}
    K_0/K&=1+\phi P(\alpha,\nu),\label{eq:Kdil}\\
    G_0/G&=1+\phi Q(\alpha, \nu).\label{eq:Gdil}
  \end{align}
\end{linenomath}
The first (and most obvious) method for evaluating \nuc{} in the fluid-saturated case would be to (1) insert the asymptotic expressions of \citet{david11} for $P$ and $Q$ in limiting cases of thin-cracks, nearly spherical pores and needles in \eqref{eq:Kdil} and \eqref{eq:Gdil} to compute the dry moduli in the limit of small porosity, (2) use Gassmann's equation to compute the saturated moduli and (3) solve for \nuc{} for each pore geometry. However, we found this approach rather cumbersome. Alternatively, the fluid-saturated Poisson ratio is readily evaluated by solving a modified DEM scheme in the limit of small porosity, using the shear compliance of dry pores $Q$ (unaffected by fluid saturation in the low frequency limit) and an effective pore bulk compliance equal to $(1-\zeta)P_\mathrm{u}(\alpha, \nu, \zeta)$, where $P_\mathrm{u}$ is the bulk compliance of fluid-saturated inclusions \citep{david12b}. This approach has been shown to be rigorously equivalent to the first method described above in the limit of small porosity \citep{david12b}. The critical Poisson's ratio is given by setting
\begin{linenomath}
  \begin{equation}\label{eq:nuundrained}
    Q(\alpha, \nu)-(1-\zeta)P_\mathrm{u}(\alpha, \nu, \zeta)=0.
  \end{equation}
\end{linenomath}

\begin{description}
\item[\textbf{Thin cracks} ($\alpha\ll1$)] Series expansions of $P_\mathrm{u}$ and $Q$ for small $\alpha$ and small $\zeta$ (in that order) are used, and yield the following approximation:
\begin{linenomath}
  \begin{equation}
    \nu_\mathrm{0,crit} \sim \frac{40\zeta}{81\pi\alpha}.
  \end{equation}
\end{linenomath}
Because of the order in which the series expansions are performed, this approximation is valid for $\alpha\ll\zeta$. We did not find any useful approximation for the case $\zeta\leq\alpha\ll1$.
\item[\textbf{Spheres} ($\alpha\sim1$)] Series expansions near $\alpha=1$ and small $\zeta$ result in an approximation that is independent from $\alpha$ (at least to first order):
\begin{linenomath}
  \begin{equation}
    \nu_\mathrm{0,crit}\sim \frac{1}{5}\left(1+4\zeta\right).
  \end{equation}
\end{linenomath}

\item[\textbf{Needles} ($\alpha\gg1$)] Series expansions for large $\alpha$ and small $\zeta$ yield
\begin{linenomath}
  \begin{equation}
    \nu_\mathrm{0,crit}\sim \frac{1}{8}\left(7-\sqrt{29}\right) + \frac{203+36\sqrt{29}}{522}\zeta.
  \end{equation}
\end{linenomath}
\end{description}

\section{Asymptotic forms in the saturated, unrelaxed case}
\label{ax:unrelaxed}

In the previous Section we derived asymptotes for the undrained saturated case, which is given by inserting the dry moduli from the DEM scheme into Gassmann's equations. The undrained case corresponds to the low frequency limit \citep{leravalec96}, where the fluid pressure is the same in all pores within a representative elementary volume over which the averaging procedure is performed. In the high frequency limit, also called ``unrelaxed'' or ``saturated-isolated'' limit, the fluid pressure is not equilibrated between each pore. This is results in saturated efective moduli that are equal or higher than those predicted in the undrained, low frequency regime \citep{leravalec96}. Although this case is commonly not directly relevant to the low frequencies used in conventional seismology \citep{li18}, for completeness we include here the key asymptotes for \nuc{} for each aspect ratio limit. Although the evolution of $\nu$ with increasing $\phi$ is quantitatively different compared to the undrained, low frequency case, only minor differences are found for the critical Poisson ratio \nuc{}.

The high frequency, unrelaxed critical Poisson's ratio $\nu_\mathrm{0,crit}^\mathrm{HF}$ is given by setting
\begin{linenomath}
  \begin{equation}\label{eq:nuunrelaxed}
    Q_\mathrm{u}(\alpha, \nu, \zeta)-(1-\zeta)P_\mathrm{u}(\alpha, \nu, \zeta)=0,
  \end{equation}
\end{linenomath}
where $Q_\mathrm{u}$ is the shear compliance of a fluid-saturated spheroidal inclusion \citep[see][for complete expression]{david12b}.

\begin{description}
  \item[\textbf{Thin cracks} ($\alpha\ll1$)] Series expansions of $P_\mathrm{u}$ and $Q_\mathrm{u}$ for small $\alpha$ and small $\zeta$ (in that order) are used, and yield the following approximation:
\begin{linenomath}
  \begin{equation}
    \nu_\mathrm{0,crit}^\mathrm{HF} \sim \frac{8\zeta}{27\pi\alpha} \simeq 0.094\,\zeta/\alpha.
  \end{equation}
\end{linenomath}

\item[\textbf{Spheres} ($\alpha\sim1$)] This case is rigorously equivalent to the undrained case:
\begin{linenomath}
  \begin{equation}
    \nu_\mathrm{0,crit}^\mathrm{HF}\sim \frac{1}{5}\left(1+4\zeta\right).
  \end{equation}
\end{linenomath}

\item[\textbf{Needles} ($\alpha\gg1$)] Series expansions for large $\alpha$ and small $\zeta$ yield
\begin{linenomath}
  \begin{equation}
    \nu_\mathrm{0,crit}^\mathrm{HF}\sim \frac{1}{8}\left(7-\sqrt{29}\right) + \frac{551+91\sqrt{29}}{1392}\zeta \simeq 0.202 + 0.748\,\zeta.
  \end{equation}
\end{linenomath}
\end{description}

\balance

\end{document}